\documentclass[sn-mathphys,Numbered]{sn-jnl}
\usepackage{graphicx}%
\usepackage{multirow}%
\usepackage{amsmath,amssymb,amsfonts}%
\usepackage{amsthm}%
\usepackage{mathrsfs}%
\usepackage[title]{appendix}%
\usepackage{xcolor}%
\usepackage{textcomp}%
\usepackage{manyfoot}%
\usepackage{booktabs}%
\usepackage{algorithm}%
\usepackage{algorithmicx}%
\usepackage{algpseudocode}%
\usepackage{listings}%

\begin{document}

\title{Evading Quantum Mechanics \`{a} la Sudarshan: quantum-mechanics-free subsystem as a realization of Koopman-von Neumann mechanics}  

\author*[1,2]{\fnm{Zurab K.} \sur{Silagadze}} \email{silagadze@inp.nsk.su}

\affil[1]{\orgname{Budker Institute of Nuclear Physics}, \orgaddress{\postcode{630 090}, \city{Novosibirsk}, \country{Russia}}}
\affil[2]{\orgname{Novosibirsk State University}, \orgaddress{\postcode{630 090}, \city{Novosibirsk}, \country{Russia}}}

\abstract{Tsang and Caves suggested the idea of a quantum-mechanics-free subsystem in 2012. We contend that Sudarshan's viewpoint on Koopman-von Neumann mechanics is realized in the quantum-mechanics-free subsystem. Since quantum-mechanics-free subsystems are being experimentally realized, Koopman-von Neumann mechanics is essentially transformed into an engineering science.}

\keywords{Koopman-von Neumann mechanics; Quantum-mechanics-free subsystem} 

\maketitle

\abstract{Tsang and Caves suggested the idea of a quantum-mechanics-free subsystem in 2012. We contend that Sudarshan's viewpoint on Koopman-von Neumann mechanics is realized in the quantum-mechanics-free subsystem. Since quantum-mechanics-free subsystems are being experimentally realized, Koopman-von Neumann mechanics is essentially transformed into an engineering science.}


The notion of quantum-mechanics-free subsystem (QMFS) was introduced by Tsang and Caves in \cite{Tsang_2012}. They showed that it is possible to design the dynamics of coupled quantum systems in such a way that one can construct a subsystem in which a given classical dynamics is realized, thus avoiding the measurement back-action of quantum mechanics. Although they mention that the Heisenberg picture equations of motion for QMFS was first proposed by Koopman \cite{Koopman_1931} as a formulation of classical Hamiltonian dynamics in Hilbert space, in fact the connection between QMFS and Koopman-von Neumann mechanics \cite{Koopman_1931,Neumann_1932} is deeper. Namely, quantum-mechanics-free subsystem is nothing but the realization of Sudarshan's perspective \cite{Sudarshan_1976,Sudarshan_1979} on the Koopman-von Neumann mechanics.

The idea of QMFS can be explained as follows \cite{Tsang_2012}. Let us assume that the Hamiltonian of a quantum system is equal to (or to its symmetrized version with respect to non-commutative variables)
\begin{equation}
    H=f(q,p,t)P+g(q,p,t)Q+h(q,p,t),
    \label{eq1}
\end{equation}
where $f(q,p,t)$, $g(q,p,t)$, $h(q,p,t)$ are arbitrary functions, and $q,P$, $Q,p$ are two pairs of quantum-mechanical conjugate variables subject to canonical commutation relations
\begin{equation}
      [q,P]=i\hbar,\;\;[Q,p]=i\hbar,
      \label{eq2}
\end{equation}
with other commutators between them equal to zero. Then the Heisenberg picture equations of motion for commuting variables $q,p$ 
\begin{equation}
    \frac{dq}{dt}=\frac{\partial H}{\partial P}=f(q,p,t),\;\;\;
    \frac{dp}{dt}=-\frac{\partial H}{\partial Q}=-g(q,p,t),
    \label{eq2A}
\end{equation}
do not contain "hidden" variables $Q,P$ and will correspond to the classical Hamiltonian dynamics if there exists the classical Hamiltonian function $H_{cl}(q,p,t)$ such that 
\begin{equation}
    f(q,p,t)=\frac{\partial H_{cl}}{\partial p},\;\;\;
    g(q,p,t)=\frac{\partial H_{cl}}{\partial q}.
    \label{eq3}
\end{equation}
A non-trivial aspect of the QMFS proposal is that the quantum dynamics described by the Hamiltonian (\ref{eq1}) can actually be realized. According to \cite{Tsang_2012}, a pairing of positive- and negative-mass oscillators can be used for this goal.\footnote{I am tempted to recall here the maxim quoted in \cite{Zhang:2021} and attributed to Sidney Coleman: ``The career of a young theoretical physicist consists of treating the harmonic oscillator in ever-increasing levels of abstraction."} Indeed, the quantum Hamiltonian in this case is 
\begin{equation}
    H=\frac{p_1^2}{2m}+\frac{1}{2}m\omega^2q_1^2-\frac{p_2^2}{2m}-\frac{1}{2}m\omega^2q_2^2.
    \label{eq4}
\end{equation}
In terms of new  canonical variables
\begin{equation}
    q=q_1+q_2,\;\;Q=\frac{1}{2}(q_1-q_2),\;\;p=p_1-p_2,\;\;P=\frac{1}{2}(p_1+p_2),
    \label{eq5}
\end{equation}
the Hamiltonian (\ref{eq4}) takes the form
\begin{equation}
    H=\frac{pP}{m}+m\omega^2qQ,
    \label{eq6}
\end{equation}
and is exactly of the type (\ref{eq1}) and (\ref{eq3}) with $H_{cl}=\frac{p^2}{2m}+\frac{1}{2}m\omega^2q^2$.

The concept of a negative mass oscillator assumes that the entire Hamiltonian is inverted, not just the potential. Such Hamiltonians are unusual in quantum mechanics because they are unbounded from below and therefore indicate inherent instability. Nevertheless, the description in terms of a negative mass harmonic oscillator is applicable to collective excitations in various nonequilibrium systems, and the corresponding negative mass instability has been observed experimentally \cite{Kohler_2018}. Also note that (\ref{eq4}) may be a low-energy effective description of a more realistic situation. For example, if we add $\frac{1}{2\Lambda}\left (\frac{p_2^2}{2m}+\frac{1}{2}m\omega^2q_2^2\right)^2$ term to (\ref{eq4}), the resulting Hamiltonian will differ negligibly from the original one at energies $E\ll\Lambda$, but is bounded from below \cite{Gross:2020tph}.

An oscillator with an effective negative mass can be built, for example, using ensembles of atomic spins \cite{Moller_2017}, and several experimental works have demonstrated the feasibility of creating QMFSs \cite{Moller_2017,Mercier_de_L_pinay_2021,Ockeloen-Korppi_2016}.

To the best of our knowledge, the QMFS literature fails to recognize the fact that QMFS is essentially nothing more than an implementation of the Koopman-von Neumann (KvN) mechanics. The same applies to literature related to KvN mechanics. Perhaps this failure is explained by the significant difference between the metaphysics of quantum mechanics and the metaphysics of classical mechanics. In reality, decoherence, losses, and various imperfections will make it inevitable that observed QMFS will be contaminated by quantum noise due to induced couplings to ``hidden" quantum variables \cite{Tsang_2012}. It is therefore natural to think quantum mechanically in the QMFS literature. On the other hand, in Koopman-von Neumann mechanics you are not forced to accept Sudarshan's interpretation (described below) and consider the KvN system as part of a twice larger quantum system. In this case, the fact that Liouvillian is not bounded from below does not create any problems, there are no ``hidden" quantum variables, and the metaphysics is essentially classical, despite the quantum-like Hilbert space formalism.

For modern overview of the KvN mechanics, see, for example, \cite{Mauro_2003,Chashchina_2019} and references therein. However, to substantiate the above statement about QMFS, it is enough to recall the very basics of KvN mechanics. The starting point is the Liouville equation of classical statistical mechanics:
\begin{equation}
    \frac{\partial\rho(q,p,t)}{\partial t}=\frac{\partial H_{cl}}{\partial q}\frac{\partial \rho}{\partial p}-\frac{\partial H_{cl}}{\partial p}\frac{\partial \rho}{\partial q}.
    \label{eq7}
\end{equation}
Due to the linearity of this equation with respect to the derivatives $\frac{\partial \rho}{\partial p}$ and $\frac{\partial \rho}{\partial q}$, we can introduce the classical wave function $\psi(q,p,t)=\sqrt{\rho(q,p,t)}$, which obeys the same Liouville equation (\ref{eq7}), which can be rewritten in a Schr\"{o}dinger-like form
\begin{equation}
    i\frac{\partial\psi(q,p,t)}{\partial t}=L\psi,\;\;\;L=i\left(\frac{\partial H_{cl}}{\partial q}\frac{\partial }{\partial p}-\frac{\partial H_{cl}}{\partial p}\frac{\partial }{\partial q}\right ).
    \label{eq8}
\end{equation}
Based on this observation, it is possible to develop a Hilbert space formulation of classical mechanics that is completely reminiscent of the quantum formalism, except that, of course, all interference effects are absent \cite{Koopman_1931,Neumann_1932}. 

Sudarshan gave a very interesting interpretation of KvN mechanics \cite{Sudarshan_1976,Sudarshan_1979}. If we introduce $Q$ and $P$ operators as follows
\begin{equation}
    Q=i\hbar\frac{\partial }{\partial p},\;\;\;P=-i\hbar\frac{\partial }{\partial q},
    \label{eq9}
\end{equation}
then the equation (\ref{eq8}) takes the form
\begin{equation}
    i\hbar\frac{\partial\psi(q,p,t)}{\partial t}=H\psi,\;\;\;H=\frac{\partial H_{cl}}{\partial q}Q+\frac{\partial H_{cl}}{\partial p}P,
    \label{eq10}
\end{equation}
and it can be interpreted as the Schr\"{o}dinger equation in the $(q,p)$-representation (with diagonal operators $q$ and $p$) of a genuine quantum system with two pairs of canonical variables $(q,P)$ and $(Q,p)$. The similarity with the QMFS idea is obvious, since the quantum Hamiltonian in the equation (\ref{eq10}) has exactly the same type as given by (\ref{eq1}) and (\ref{eq3}), and hence $(q,p)$ subsystem is nothing but QMFS.

The introduction of Planck's constant in (\ref{eq9}) and (\ref{eq10}) is quite formal, just to remind that KvN mechanics is not a derivative of quantum mechanics in the $\hbar\to 0$ limit. The limit $\hbar\to 0$ means something more than just a criterion for the commutativity of the position and momentum operators in the classical limit, since $\hbar$ is present not only in the canonical commutation relation, but also as a factor in front of the time derivative in the Schr\"{o}dinger equation. KvN mechanics can be motivated by another type of limit based on the quasi-probabilistic Wigner distribution function \cite{Bondar_2012,Sen:2022vig}.

The absence of interference effects in Sudarshan's interpretation is guaranteed by the superselection principle, according to which any operator containing hidden quantum variables $Q$ and $P$ is not a classical observable. Such superselection principles that ensure classicality can arise naturally in the course of a Hamiltonian evolution which correlates one quantum system with another quantum systems \cite{Zurek:1982ii}. Unlike quantum mechanics, where the eigenvalues of superselection operators are constants of motion, in KvN mechanics, the quantum Hamiltonian in (\ref{eq10}) (the Liouville operator disguised from a classical point of view) is not a classical observable and can cause transitions from one eigenspace of superselection operators $q$ and $p$ into other eigenspaces, since it does not commute with these operators.

I think that the identity of quantum-mechanics-free subsystems with Sudarshan's interpretation of KvN mechanics, combined with the fact that such systems were actually implemented experimentally, is of great importance for Koopman-von Neumann mechanics, as it makes KvN mechanics, in a sense, engineering science.  

The renewed interest in KvN mechanics was caused by the need to build a suitable framework for hybrid classical-quantum systems (see, for example, \cite{Bondar_2023,Morgan_2020} and references therein). I hope that the realization of the fact that quantum-mechanics-free subsystems are described by KvN mechanics will boost this interest in KvN mechanics.

There is one more curious aspect of QMFS-KvN mechanics connection. In \cite{Chashchina_2019}, it was suggested that the modification of quantum mechanics supposedly expected from quantum gravity could lead to a deformation of classical mechanics, effectively destroying classicality, if Sudarshan's views on KvN mechanics are taken seriously. In \cite{Chashchina_2019} this was of purely academic interest, since you are not required to accept the Sudarshan interpretation in order to develop the KvN mechanics. However, we now see that the existence of quantum-mechanics-free subsystems indicates that we should take Sudarshan's interpretation of KvN mechanics seriously. Therefore, we expect that, due to the universal nature of gravity, if the effects of quantum gravity do modify quantum mechanics, these effects will destroy the classical dynamics in QMFS, since the hidden variables will no longer be hidden from the classical subspace. It may be worth looking for such effects in quantum-mechanics-free subsystems.

\bibliography{KvN_Eng}

\end{document}